\documentclass[aps,pra,twocolumn,superscriptaddress,floatfix]{revtex4-2}

\usepackage{amsmath,amssymb,amsthm}
\usepackage{bm}
\usepackage{graphicx}
\usepackage[T1]{fontenc}
\usepackage[utf8]{inputenc}
\usepackage{quantikz}
\usepackage[compat=1.1.0]{tikz-feynman}
\usepackage{slashed}
\usepackage{xcolor}
\usepackage{hyperref}

\DeclareMathOperator{\diag}{diag}

\begin{document}

\title{Quantum circuit representation and incomplete Bell measurements in Compton scattering}

\author{Wei Xie}
\email{xiewei@ctgu.edu.cn}
\affiliation{Department of Physics, Liaoning Normal University, No.~850 Huanghe Road, Dalian 116029, China}
\affiliation{College of Mathematics and Physics, China Three Gorges University, No.~8 University Road, Yichang 443002, China}
\affiliation{Center for Astronomy and Space Sciences, China Three Gorges University, No.~8 University Road, Yichang 443002, China}

\author{Yu-Chen Guo}
\email{ycguo@lnnu.edu.cn}
\affiliation{Department of Physics, Liaoning Normal University, No.~850 Huanghe Road, Dalian 116029, China}
\affiliation{Center for Theoretical and Experimental High Energy Physics, Liaoning Normal University, No.~850 Huanghe Road, Dalian 116029, China}

\author{Shuang Li}
\email{lish@ctgu.edu.cn}
\affiliation{College of Mathematics and Physics, China Three Gorges University, No.~8 University Road, Yichang 443002, China}
\affiliation{Center for Astronomy and Space Sciences, China Three Gorges University, No.~8 University Road, Yichang 443002, China}

\author{Ji-Chong Yang}
\email{yangjichong@lnnu.edu.cn(Corresponding author)}
\affiliation{Department of Physics, Liaoning Normal University, No.~850 Huanghe Road, Dalian 116029, China}
\affiliation{Center for Theoretical and Experimental High Energy Physics, Liaoning Normal University, No.~850 Huanghe Road, Dalian 116029, China}

\date{\today}

\begin{abstract}
We formulate massless tree-level Compton scattering as an effective two-qubit
operation on electron and photon helicities. In the Bell basis, the scattering
map separates into two parity sectors and admits a simple quantum circuit
representation with one nonunitary filter. After Kraus normalization, this
filter defines an unsharp incomplete Bell measurement that distinguishes the
two Bell-parity sectors but not the relative Bell label within each sector. We
further show that the incomplete measurement alone does not surpass the
classical teleportation threshold, whereas a subsequent complete Bell-basis
readout enables unit conditional teleportation fidelity.
\end{abstract}

\maketitle

\section{Introduction}

Quantum information and computation have become increasingly 
relevant to high energy physics \cite{QIHEP,Barr:2024djo,Bertlmann2006}.
Studies range from Bell inequalities, coherence, and entanglement in
neutral-meson systems
\cite{Bertlmann:2001ea,DiDomenico:1995ky,DiDomenico:2007zza,Dixit:2018gjc}
to quantum correlations in neutrino oscillations
\cite{Banerjee:2015mha,Bittencourt:2022tcl,Bittencourt:2023asd,
Blasone:2021cau,Blasone:2023qqf,Blasone:2007vw,Formaggio:2016cuh}.
At high energy colliders, entanglement and Bell correlations
 have been explored in top-quark, Higgs, vector-boson, and other
high-energy processes
\cite{Afik:2022kwm,Afik:2023dgh,Fabbrichesi:2021npl,LHC,CMS,
HZZ,Barr:2021zcp,ATLAS:2026hye,Gao:2025kdi,Jaloum:2025bkx}.
More broadly, entanglement has been investigated as a probe of symmetry
structure and possible physics beyond the Standard Model
\cite{Beane:2018oxh,Carena:2025wyh,Duch:2024pwm,Chang:2024wrx}.

Particle scattering processes provide a natural
setting in which quantum information ideas can be implemented.
Spin or helicity degrees of freedom define finite dimensional quantum
states, while scattering amplitudes determine their transformations.
 {Building on the helicity formalism~\cite{JacobWick}, a substantial literature has
investigated entanglement generation and redistribution in tree-level QED 
processes~\cite{Cervera-Lierta:2017tdt,Fan,Fan:2017mth,Serafini,
Fedida:2024dwc,Fonseca:2021uhd,Kowalska:2024kbs,
BlasoneSE2,NoiChaos,Blasone:2024dud,Cao:2026ljz,Cao:2026mza}.
Related low-energy studies include work on $e^+e^-$ scattering and measurements
of Bhabha-to-M{\o}ller cross-section ratios~\cite{Scherdin1991,Goebel1993}.
Other work has connected scattering entanglement with cross sections,
quantum parameter estimation, and experimentally accessible particle and
photon correlations~\cite{Araujo,Asenov:2025ueu,Gao:2025kdi,Clarke2026}.

Scattering amplitudes are conventionally expressed as complex numbers labeled by
the quantum numbers of the incoming and outgoing particles. At fixed external
momenta, however, the full set of amplitudes naturally defines a linear map from
the internal space of the incoming state to that of the outgoing state.
When the relevant internal degrees of freedom are two-level helicity states, this
map acts directly on qubits, providing a possible bridge between particle scattering 
and quantum computation. This perspective makes it possible to characterize a
scattering process in terms of quantum circuits. Along this line, quantum 
circuits have been designed to generate maximally entangled states~\cite{Cervera-Lierta:2019juk}. 

More recently, fixed-kinematics helicity transition
matrices for lepton scattering processes have been decomposed explicitly into
unitary gates and nonunitary singular value factors~\cite{Xie:2026ixr}. 
These developments establish the
amplitude-as-channel viewpoint and motivate a more detailed analysis of the
measurement structure and quantum-information content encoded in scattering
amplitudes. The aim of the present work is to provide an analytic realization of
this program for Compton scattering and to determine precisely what Bell-state
information is encoded in the nonunitary part of the corresponding helicity
map. 

The massless Compton
helicity matrix takes an especially simple form, which is diagonal in the
electron-photon helicity basis and reduces, in the Bell basis, to two
diagonal parity-sector blocks. This structure allows an exact two-qubit circuit
representation and cleanly separates the unitary basis transformations from
the nonunitary component. The nonunitary part
defines a binary quantum measurement whose outcome probabilities depend only
on the Bell-parity sector. The resulting measurement is
therefore an incomplete and, at a generic scattering angle, unsharp Bell
measurement, which provides only statistical information about Bell parity rather
than a deterministic Bell bit.

We also separate two logically different teleportation questions. First, we
determine the best fidelity obtainable if only the binary outcome of the
incomplete measurement is retained. Second, we consider an assisted heralded
scheme in which the selected Compton Kraus branch is followed by a complete
Bell-basis readout. The latter recovers unit fidelity because the selected
branch maps the Bell basis bijectively at every interior scattering angle.

The remainder of this paper is organized as follows.
Section~\ref{sec:amplitude} presents the tree-level Compton scattering
amplitude and constructs the corresponding helicity transition matrix.
Section~\ref{sec:bell} reformulates the transition matrix in the Bell basis
and introduces the normalized incomplete measurement.
Section~\ref{sec:circuit} develops the associated quantum-circuit
representation.
Section~\ref{sec:teleportation} examines the implications for the two
teleportation scenarios considered in this work.
Finally, Section~\ref{sec:conclusion} summarizes the main results.

\section{Compton scattering amplitude and its helicity transition matrix}
\label{sec:amplitude}

At tree level, Compton scattering receives the two electron-exchange
contributions shown in Fig.~\ref{fig:diagrams}. 
The incoming electron and photon carry
four-momenta $p_1$ and $p_2$ and helicities $\lambda_1$ and $\lambda_2$; the
outgoing electron and photon carry $p_3,p_4$ and $\lambda_3,\lambda_4$. Electron
helicities are labeled by $\lambda_1,\lambda_3\in\{+,-\}$ and photon
helicities by $\lambda_2,\lambda_4\in\{+,-\}$. 
With $\slashed a\equiv\gamma_\mu a^\mu$ and electron mass $m$, the invariant
amplitude is
\begin{align}
\mathcal M_{\lambda_3\lambda_4;\lambda_1\lambda_2}
=-ie^2\,\bar u_3
\bigg[&
\slashed\epsilon_4^{\,*}
\frac{\slashed p_1+\slashed p_2+m}{s-m^2}
\slashed\epsilon_2
\nonumber\\
&+
\slashed\epsilon_2
\frac{\slashed p_1-\slashed p_4+m}{u-m^2}
\slashed\epsilon_4^{\,*}
\bigg]u_1 .
\label{eq:amplitude}
\end{align}
Here
$u_1=u(p_1,\lambda_1)$,
$\bar u_3=\bar u(p_3,\lambda_3)$,
$\epsilon_2=\epsilon(p_2,\lambda_2)$, and
$\epsilon_4=\epsilon(p_4,\lambda_4)$. 
The explicit external-state conventions used to fix the helicity phases are
given in Appendix~\ref{app:conventions}.

\begin{figure*}[t]
\centering
\begin{tikzpicture}
\begin{feynman}
\vertex (a1) at (-3,1.5) {$e^-(p_1)$};
\vertex (a2) at (-3,-1.5) {$\gamma(p_2)$};
\vertex (v1) at (-1,0);
\vertex (v2) at (1,0);
\vertex (a3) at (3,1.5) {$e^-(p_3)$};
\vertex (a4) at (3,-1.5) {$\gamma(p_4)$};
\diagram*{
(a1) -- [fermion] (v1),
(a2) -- [photon] (v1),
(v1) -- [fermion, edge label=$p_1+p_2$] (v2),
(v2) -- [fermion] (a3),
(v2) -- [photon] (a4)
};
\end{feynman}
\node at (0,-2.15) {(a) $s$ channel};
\end{tikzpicture}
\hspace{1.0cm}
\begin{tikzpicture}
\begin{feynman}
\vertex (b1) at (-3,1.5) {$e^-(p_1)$};
\vertex (b2) at (-3,-1.5) {$\gamma(p_2)$};
\vertex (w1) at (-1,0);
\vertex (w2) at (1,0);
\vertex (b3) at (3,1.5) {$e^-(p_3)$};
\vertex (b4) at (3,-1.5) {$\gamma(p_4)$};
\diagram*{
(b1) -- [fermion] (w1),
(w1) -- [photon] (b4),
(w1) -- [fermion, edge label=$p_1-p_4$] (w2),
(b2) -- [photon] (w2),
(w2) -- [fermion] (b3)
};
\end{feynman}
\node at (0,-2.15) {(b) $u$ channel};
\end{tikzpicture}
\caption{Tree-level Compton-scattering diagrams. The two contributions differ
by the ordering of the photon vertices on the electron line. Their sum is
required by electromagnetic gauge invariance.}
\label{fig:diagrams}
\end{figure*}
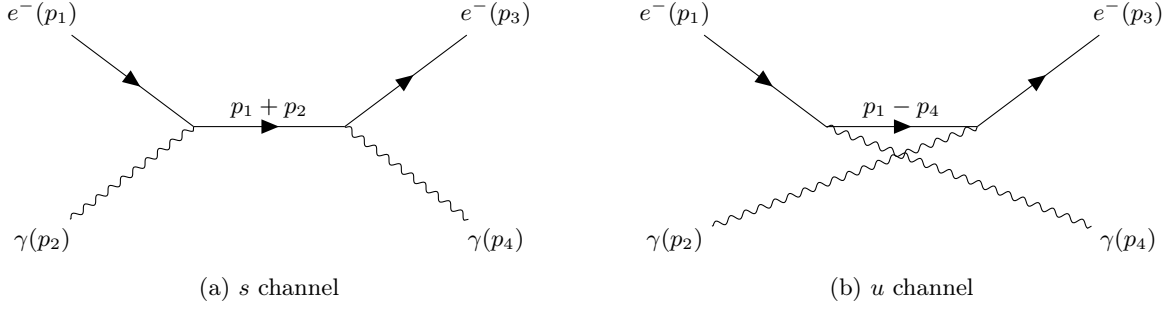

We take the massless limit $m=0$ and work in the center-of-mass frame. 
The Mandelstam variables are
\begin{equation}
\begin{aligned}
s&=(p_1+p_2)^2=4E^2,\\
t&=(p_1-p_3)^2=-2E^2(1-\cos\theta),\\
u&=(p_1-p_4)^2=-2E^2(1+\cos\theta).
\end{aligned}
\label{eq:stu}
\end{equation}

The helicity transition matrix is defined by~\cite{Xie:2026ixr}
\begin{equation}
T=
\begin{pmatrix}
\mathcal{M}_{++;++} &
\mathcal{M}_{++;+-} &
\mathcal{M}_{++;-+} &
\mathcal{M}_{++;--}
\\
\mathcal{M}_{+-;++} &
\mathcal{M}_{+-;+-} &
\mathcal{M}_{+-;-+} &
\mathcal{M}_{+-;--}
\\
\mathcal{M}_{-+;++} &
\mathcal{M}_{-+;+-} &
\mathcal{M}_{-+;-+} &
\mathcal{M}_{-+;--}
\\
\mathcal{M}_{--;++} &
\mathcal{M}_{--;+-} &
\mathcal{M}_{--; -+} &
\mathcal{M}_{--;--}
\end{pmatrix}.
\end{equation}
Rows label the outgoing electron-photon helicities and columns label the
incoming helicities. In the computational basis we identify
\begin{equation}
\begin{aligned}
|0\rangle_e&\equiv|+\rangle_e,
&|1\rangle_e&\equiv|-\rangle_e,\\
|0\rangle_\gamma&\equiv|+\rangle_\gamma,
&|1\rangle_\gamma&\equiv|-\rangle_\gamma.
\end{aligned}
\label{eq:qubitencoding}
\end{equation}
and order the two-qubit states as
$|++\rangle,|+-\rangle,|-+\rangle,|--\rangle$.
Throughout the two-qubit discussion, an unsubscripted $I$ denotes the
$2\times2$ identity acting on one qubit; tensor products specify which factor
of the electron-photon Hilbert space it acts upon.

Massless QED conserves electron chirality, and hence electron helicity, along
the fermion line. Direct evaluation of 
Eq.~\eqref{eq:amplitude} shows that the photon-helicity-changing entries also cancel. 
The two independent amplitude coefficients are
\begin{align}
A(\theta)
&\equiv\mathcal M_{++;++}=-\frac{2\sqrt2\,i e^2}{\sqrt{1+\cos\theta}},
\nonumber \\
B(\theta)
&\equiv\mathcal M_{+-;+-}
=i\sqrt2\,e^2\sqrt{1+\cos\theta}.
\label{eq:AB}
\end{align}
These expressions are understood away from the backward endpoint
$\theta=\pi$, where $u\to0$ and the massless amplitude becomes singular.
The nonzero entries are
$\mathcal M_{++;++}=-A$, $\mathcal M_{+-;+-}=B$,
$\mathcal M_{-+;-+}=B$, and $\mathcal M_{--;--}=-A$. Thus
\begin{equation}
T(\theta)=
\begin{pmatrix}
-A & 0 & 0 & 0\\
0 & B & 0 & 0\\
0 & 0 & B & 0\\
0 & 0 & 0 & -A
\end{pmatrix}.
\label{eq:Tmatrix}
\end{equation}

No dependence on the beam energy appears in
Eq.~\eqref{eq:AB}. In the massless theory, the QED coupling is dimensionless
and no additional mass scale is present. At fixed scattering angle, the
amplitude can therefore depend only on dimensionless ratios of Mandelstam
invariants. The ratio of two
nonzero amplitudes is
\begin{equation}
\frac{|A|}{|B|}=\frac{2}{1+\cos\theta}.
\label{eq:ratio}
\end{equation}

As a check, averaging over the two helicity states of each incoming particle and summing 
over the final helicities yields
\begin{align}
\overline{|\mathcal M|^2}
&\equiv
\frac14
\sum_{\lambda_1,\lambda_2,\lambda_3,\lambda_4}
|\mathcal M_{\lambda_3\lambda_4;\lambda_1\lambda_2}|^2
\nonumber\\
&=
\frac14\left(2|A|^2+2|B|^2\right)
=
\frac12\left(|A|^2+|B|^2\right)
\nonumber\\
&=
2e^4\left(-\frac{u}{s}-\frac{s}{u}\right)
\nonumber\\
&=
e^4\left(
1+\cos\theta+\frac{4}{1+\cos\theta}
\right).
\label{eq:spinaverage}
\end{align}
At $\theta=\pi/2$, Eq.~\eqref{eq:spinaverage} gives
$\overline{|\mathcal M|^2}=5e^4$.

\section{Bell basis and incomplete Bell measurement}
\label{sec:bell}

\subsection{Bell-basis representation}

The four Bell states are
\begin{align}
|\Phi^\pm\rangle&=\frac{|++\rangle\pm|--\rangle}{\sqrt2},
\nonumber\\
|\Psi^\pm\rangle&=\frac{|+-\rangle\pm|-+\rangle}{\sqrt2}.
\label{eq:bellstates}
\end{align}
We order them as
$|\Phi^+\rangle,|\Phi^-\rangle,|\Psi^+\rangle,|\Psi^-\rangle$. The unitary
matrix that maps Bell-basis coordinates to computational-basis coordinates,
$|\psi\rangle_{\mathrm{comp}}=W|\psi\rangle_{\mathrm B}$, is
\begin{equation}
W=\frac{1}{\sqrt2}
\begin{pmatrix}
1&1&0&0\\
0&0&1&1\\
0&0&1&-1\\
1&-1&0&0
\end{pmatrix}.
\label{eq:W}
\end{equation}
Let $H$ be the Hadamard gate, and let
$\mathrm{CNOT}_{a\to b}$ denote a controlled-NOT gate with control qubit $a$
and target qubit $b$. With qubit $0$ representing the electron and qubit $1$
the photon,
\begin{equation}
W=\mathrm{CNOT}_{1\to0}\,
\mathrm{CNOT}_{0\to1}\,(I\otimes H).
\label{eq:Wgates}
\end{equation}

Transforming Eq.~\eqref{eq:Tmatrix} gives
\begin{equation}
T_{\mathrm B}\equiv W^\dagger T W
=\begin{pmatrix}
-A I_2&0\\0& B I_2
\end{pmatrix},
\label{eq:Tbell}
\end{equation}
where $I_2$ is the identity on each Bell-parity subspace. Therefore
\begin{align}
T_{\mathrm B}|\Phi^+\rangle&=-A|\Phi^+\rangle,&
T_{\mathrm B}|\Phi^-\rangle&=-A|\Phi^-\rangle,
\nonumber\\
T_{\mathrm B}|\Psi^+\rangle&=B|\Psi^+\rangle,&
T_{\mathrm B}|\Psi^-\rangle&=B|\Psi^-\rangle.
\label{eq:bellaction}
\end{align}
The process preserves each Bell state up to its sector-dependent amplitude.

Define the Bell-parity projectors
\begin{align}
P_\Phi&=|\Phi^+\rangle\langle\Phi^+|
+|\Phi^-\rangle\langle\Phi^-|,
\nonumber\\
P_\Psi&=|\Psi^+\rangle\langle\Psi^+|
+|\Psi^-\rangle\langle\Psi^-|.
\label{eq:projectors}
\end{align}
The singular values of $T$ are
$|A|,|A|,|B|,|B|$, and
\begin{equation}
T^\dagger T=|A|^2P_\Phi+|B|^2P_\Psi.
\label{eq:TdaggerT}
\end{equation}

\subsection{Kraus normalization and POVM}

\begin{figure*}[tb]
    \centering
    \includegraphics[width=0.8\textwidth]{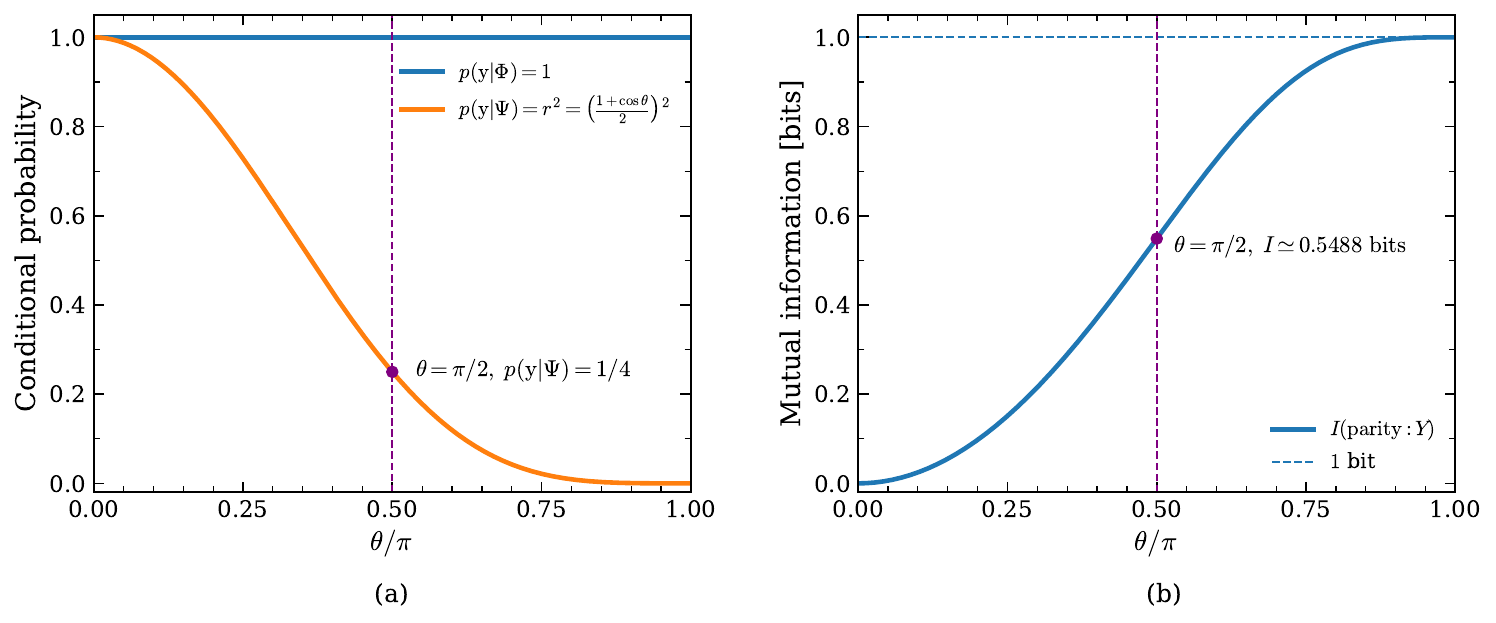}
    \caption{Angular dependence of the incomplete Bell measurement.
(a) Conditional selected-outcome probabilities 
$p(\mathrm{y}|\Phi)=1$ and $p(\mathrm{y}|\Psi)=r^2=[(1+\cos\theta)/2]^2$.
(b) Mutual information $I(\mathrm{parity}:Y)=h_2[(1+r^2)/2]-\tfrac{1}{2}h_2(r^2)$ for equal 
prior probabilities of the $\Phi$- and $\Psi$-parity sectors. 
The backward endpoint is understood as the normalized limit $\theta\to\pi$.}
\label{fig:incomplete_bell_measurement}
\end{figure*}

The transition amplitude $T$ is not itself a Kraus operator because its overall
normalization is not constrained by $T^\dagger T\leq I$. For any positive
constant $\alpha$ satisfying
\begin{equation}
0<\alpha\leq\frac{1}{\lVert T\rVert_\infty},
\label{eq:alphabound}
\end{equation}
the operator $\alpha T$ is a valid trace-nonincreasing quantum operation. Here
$\lVert T\rVert_\infty$ is the largest singular value of $T$. Because
$|A|\geq|B|$ for $0\leq\theta<\pi$, the maximal contraction is
\begin{equation}
K_{\mathrm y}=\frac{T}{\lVert T\rVert_\infty}=\frac{T}{|A|}.
\label{eq:Kyes}
\end{equation}
This choice maximizes the selected-branch probability within the effective
Kraus representation. This is a normalization convention rather than an additional dynamical assumption. 
 Choosing a smaller $\alpha$ would rescale the branch
probability but would not alter the normalized conditional state.

Introduce the dimensionless ratio
\begin{equation}
r(\theta)\equiv\frac{|B|}{|A|}=\frac{1+\cos\theta}{2},
\qquad 0<r\leq1,
\label{eq:r}
\end{equation}
where the strict lower bound applies at interior angles and $r\to0$ in the
formal backward limit. The selected-branch effect is
\begin{equation}
E_{\mathrm y}\equiv K_{\mathrm y}^\dagger K_{\mathrm y}
=P_\Phi+r^2P_\Psi.
\label{eq:Eyes}
\end{equation}
The complementary effect is
\begin{equation}
E_{\mathrm n}=I-E_{\mathrm y}=(1-r^2)P_\Psi.
\label{eq:Eno}
\end{equation}
The subscripts $\mathrm y$ and $\mathrm n$ label, respectively, the selected
``yes'' outcome and its complementary ``no'' outcome. Here and below, the
unsubscripted identity in a POVM completeness relation is the identity on the
four-dimensional electron-photon helicity space.
To write a complete instrument while keeping the unsuccessful branch distinct,
let $J_{\mathrm f}$ be an isometric embedding into an orthogonal failure
subspace, $J_{\mathrm f}^\dagger J_{\mathrm f}=I$, and define
\begin{equation}
K_{\mathrm n}=J_{\mathrm f}\sqrt{E_{\mathrm n}}.
\label{eq:Kno}
\end{equation}
Then
$K_{\mathrm y}^\dagger K_{\mathrm y}
+K_{\mathrm n}^\dagger K_{\mathrm n}=I$.

For a Bell-parity eigenstate, the selected-outcome probabilities are
\begin{equation}
p(\mathrm y|\Phi)=1,\qquad
p(\mathrm y|\Psi)=r^2.
\label{eq:conditionalprobabilities}
\end{equation}
Both POVM effects are proportional to the identity inside each two-dimensional
Bell-parity subspace. The measurement is therefore exactly insensitive to the
$+/-$ label. It is an incomplete Bell measurement because it probes only the
$\Phi/\Psi$ parity degree of freedom. It is also unsharp at a generic angle:
the selected outcome can arise from either parity, so one event does not
determine the parity exactly.

This distinction can be quantified. Suppose the two parities have equal prior
probabilities, and let $Y\in\{\mathrm y,\mathrm n\}$ denote the binary outcome.
Define the binary entropy
\begin{equation}
h_2(q)=-q\log_2q-(1-q)\log_2(1-q).
\label{eq:binaryentropy}
\end{equation}
The mutual information between the parity random variable and $Y$ is
\begin{equation}
I(\mathrm{parity}:Y)
=h_2\!\left(\frac{1+r^2}{2}\right)
-\frac12h_2(r^2).
\label{eq:mutualinformation}
\end{equation}

The angular dependence of the incomplete Bell measurement is shown in 
Fig.~\ref{fig:incomplete_bell_measurement}. 
As the scattering angle increases, the selected outcome becomes progressively less 
likely for the $\Psi$-parity sector, while it occurs with unit probability for the $\Phi$-parity sector. 
The binary outcome therefore becomes increasingly effective at discriminating the two 
Bell-parity sectors, as reflected by the corresponding increase of the mutual 
information $I(\mathrm{parity}:Y)$.

At $\theta=\pi/2$, $r^2=1/4$ and $I(\mathrm{parity}:Y)\simeq0.5488$ bits. 
Thus, although the measurement produces a binary record, it does not extract a 
full Bell-parity bit at this angle. In the formal normalized limit $\theta\to\pi$, $r\to0$ 
and ${E_{\mathrm y},E_{\mathrm n}}\to{P_\Phi,P_\Psi}$, so that the POVM approaches 
a sharp Bell-parity measurement. The unnormalized massless Compton amplitude itself 
is singular in this limit because $u\to0$; only the normalized POVM limit is considered here.

\section{Explicit quantum circuit representation}
\label{sec:circuit}

\begin{figure*}[tb]
\centering
\begin{quantikz}[row sep=0.55cm,column sep=0.45cm]
\lstick{$q_0$ (electron)}
&\targ{}&\ctrl{1}&\qw&\qw&\gate{R_r}&\qw&\qw&\ctrl{1}&\targ{}&\qw\\
\lstick{$q_1$ (photon)}
&\ctrl{-1}&\targ{}&\gate{H}&\qw&\qw&\qw&\gate{H}&\targ{}&\ctrl{-1}&\qw
\end{quantikz}
\caption{Quantum circuit for the maximally normalized selected Kraus operator
$K_{\mathrm y}$. The first and last three gates are the inverse Bell transform
$W^\dagger$ and Bell transform $W$, respectively. The gate
$R_r=\diag(1,r)$ is the only nonunitary element. The global phase
$i$  is omitted.}
\label{fig:circuit}
\end{figure*}
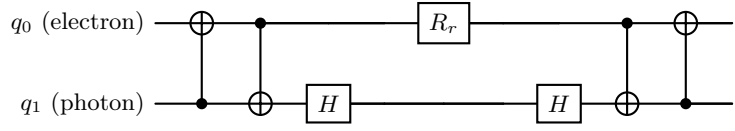

An explicit singular-value decomposition of Eq.~\eqref{eq:Tbell} is
\begin{equation}
\begin{aligned}
&T_{\mathrm B}=U\Sigma V^\dagger,\\
&U=iI,\qquad V^\dagger=I,
\end{aligned}
\label{eq:SVD}
\end{equation}
where $I$ is the $4\times4$ identity. The singular-value matrix is
\begin{equation}
\Sigma=\diag(|A|,|A|,|B|,|B|).
\label{eq:Sigma}
\end{equation}
The phase $i$ is global. Define
\begin{equation}
x(\theta)=\frac12\ln\frac{|A|}{|B|}
=\frac12\ln\frac{2}{1+\cos\theta}.
\label{eq:x}
\end{equation}
Then the unnormalized singular-value factor is
\begin{equation}
\Sigma=2e^2e^{xZ}\otimes I.
\label{eq:Sigmaexponential}
\end{equation}
The helicity transition amplitude is therefore
\begin{equation}
T=W i\Sigma W^\dagger.
\label{eq:Tcircuit}
\end{equation}

For the normalized measurement operation it is preferable to factor out the
largest singular value. Let
\begin{equation}
R_r=\diag(1,r),\qquad
D_r=R_r\otimes I=\diag(1,1,r,r).
\label{eq:Dr}
\end{equation}
Since $r=e^{-2x}$,
$D_r=e^{-x}e^{xZ}\otimes I$ and $D_r^\dagger D_r\leq I$. The selected Kraus
operator is
\begin{equation}
K_{\mathrm y}=W iD_r W^\dagger.
\label{eq:Kcircuit}
\end{equation}
All factors except $D_r$ are unitary. The single nonunitary gate $R_r$ acts on
the Bell-parity qubit: it leaves the $\Phi$ sector unchanged and attenuates the
$\Psi$ sector by $r$.

The chronological gate sequence corresponding to Eq.~\eqref{eq:Kcircuit} is
shown in Fig.~\ref{fig:circuit}. The first and last three gates implement
$W^\dagger$ and $W$, respectively, while the central gate is the Kraus filter;
the remaining factor is only the global phase $i$.

\section{Teleportation implications}
\label{sec:teleportation}

\subsection{Teleportation identity and notation}

Let qubit $1$ contain an unknown normalized state
\begin{equation}
|\psi\rangle_1=c_0|0\rangle_1+c_1|1\rangle_1,\qquad
|c_0|^2+|c_1|^2=1,
\label{eq:unknownstate}
\end{equation}
and let qubits $2$ and $3$ share the Bell state $|\Phi^+\rangle_{23}$. Qubits
$1$ and $2$ are held by Alice, and qubit $3$ by Bob. The standard teleportation
identity is
\begin{align}
|\psi\rangle_1|\Phi^+\rangle_{23}
=\frac12\big(&|\Phi^+\rangle_{12}I|\psi\rangle_3
+|\Phi^-\rangle_{12}Z|\psi\rangle_3
\nonumber\\
&+|\Psi^+\rangle_{12}X|\psi\rangle_3
+|\Psi^-\rangle_{12}XZ|\psi\rangle_3\big),
\label{eq:teleportidentity}
\end{align}
where an irrelevant phase in the last term has been absorbed into
$|\Psi^-\rangle$. Define the Bell-label set
$\mathcal B\equiv\{\Phi^+,\Phi^-,\Psi^+,\Psi^-\}$ in the displayed order.
We denote the four Bob-side Pauli operators appearing in
Eq.~\eqref{eq:teleportidentity} by $U_\beta$, where
$\beta\in\mathcal B$, so that $U_{\Phi^+}=I$, $U_{\Phi^-}=Z$,
$U_{\Psi^+}=X$, and $U_{\Psi^-}=XZ$.

\subsection{Using only the incomplete binary outcome}

Throughout this subsection, we adopt the maximal Kraus normalization defined in Eq.~\eqref{eq:Kyes}.
Before the measurement, the four Bell labels in Eq.~\eqref{eq:teleportidentity}
occur with equal probability $1/4$. The joint probabilities of the Bell label
and the selected outcome are
\begin{equation}
q_{\mathrm y}(\beta)=\frac14(1,1,r^2,r^2),
\label{eq:qyes}
\end{equation}
where the entries follow the Bell ordering used above. For the complementary
outcome,
\begin{equation}
q_{\mathrm n}(\beta)=\frac14(0,0,1-r^2,1-r^2).
\label{eq:qno}
\end{equation}
Summing these distributions gives
\begin{equation}
p_{\mathrm y}=\frac{1+r^2}{2},\qquad
p_{\mathrm n}=\frac{1-r^2}{2}.
\label{eq:outcomeprobabilities}
\end{equation}

If Alice communicates only the value $Y=\mathrm y$ or $Y=\mathrm n$ of the
binary outcome variable defined above, Bob must choose one
fixed Pauli correction for each outcome. For a Pauli channel, the entanglement
fidelity after the optimal correction is the largest conditional Pauli weight.
For the selected outcome this gives
\begin{equation}
F_{\mathrm e|\mathrm{y}}=\frac{1}{2(1+r^2)},
\label{eq:Fey}
\end{equation}
where $F_{\mathrm e|\mathrm{y}}$ denotes the conditional entanglement fidelity. The
Bloch-sphere average pure-state fidelity $\overline F$ of a qubit channel is
related to its entanglement fidelity by~\cite{NielsenChuang}
\begin{equation}
\overline F=\frac{2F_{\mathrm e}+1}{3}.
\label{eq:Faverage}
\end{equation}
Therefore
\begin{equation}
\overline F_{\mathrm y}
=\frac{2+r^2}{3(1+r^2)}.
\label{eq:Fyesconditional}
\end{equation}
The complementary outcome contains two equally likely unresolved $\Psi$
labels, so $F_{\mathrm e| \mathrm{n}}=1/2$ and
\begin{equation}
\overline F_{\mathrm n}=\frac23.
\label{eq:Fnoconditional}
\end{equation}
The optimal fidelity obtained from the incomplete record alone is consequently
\begin{align}
\overline F_{\mathrm{inc}}
&=p_{\mathrm y}\overline F_{\mathrm y}
+p_{\mathrm n}\overline F_{\mathrm n}
\nonumber\\
&=\frac23-\frac{r^2}{6}.
\label{eq:Finc}
\end{align}

\begin{figure}[tb]
\centering
\includegraphics[width=0.45\textwidth]{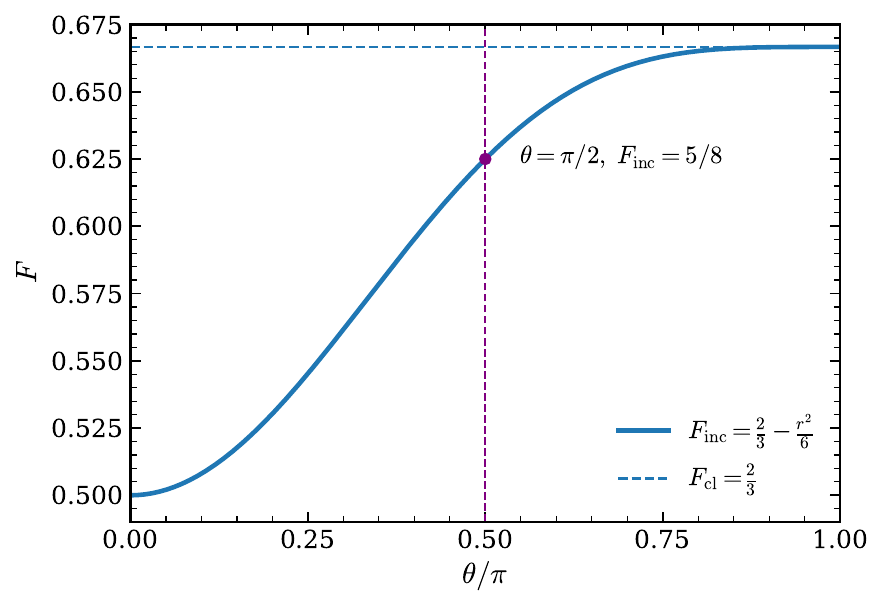}
\caption{
Angular dependence of the teleportation fidelity $F_{\rm inc}$ obtained 
when only the incomplete binary measurement record is retained.
The dashed horizontal line indicates the classical qubit-teleportation threshold $F_{\rm cl}=2/3$.
The endpoint $\theta/\pi=1$ represents the normalized backward limit $\theta\to\pi$.
}
\label{fig:teleportation_fidelity}
\end{figure}

The angular dependence of $F_{\rm inc}$ is shown in Fig.~\ref{fig:teleportation_fidelity}. 
The fidelity increases monotonically with the scattering angle.
At $\theta=0$, one has $r=1$ and hence $F_{\rm inc}=1/2$, reflecting the absence 
of Bell-parity information in the binary measurement record. At $\theta=\pi/2$, $r^2=1/4$ 
and $F_{\rm inc}=5/8$. In the formal sharp-parity limit $\theta\to\pi$, for which $r\to0$, 
the fidelity approaches $F_{\rm inc}\to2/3$. Thus, the incomplete binary measurement alone 
does not provide a quantum teleportation advantage over the classical threshold.
At $\theta=0$, $p_{\mathrm n}=0$, so the conditional fidelity for the no branch
is undefined there; the weighted fidelity $\overline F_{\mathrm{inc}}$ remains well defined.

\subsection{Assisted heralded teleportation}

We next retain the selected branch $K_{\mathrm y}$ and perform a complete
Bell-basis readout of the outgoing qubits $1$ and $2$. This readout is an
additional operation: it supplies the $+/-$ information that is absent from the
binary incomplete measurement.

From Eq.~\eqref{eq:bellaction}, the selected Compton map sends every incoming
Bell state to a unique outgoing Bell state at an interior angle. If
$\pi(\beta)$ denotes the Bell-label map
\begin{equation}
\begin{aligned}
\pi(\Phi^\pm)&=\Phi^\pm,\\
\pi(\Psi^\pm)&= \Psi^\pm.
\end{aligned}
\label{eq:permutation}
\end{equation}
then
\begin{equation}
K_{\mathrm y}|\beta\rangle=\kappa_\beta|\beta\rangle,
\label{eq:Kbellaction}
\end{equation}
with
\begin{equation}
\begin{aligned}
\kappa_{\Phi^+}=\kappa_{\Phi^-}&=i,\\
\kappa_{\Psi^+}=\kappa_{\Psi^-}&=ir.
\end{aligned}
\end{equation}
The phases follow the convention of Eq.~\eqref{eq:AB} and are irrelevant to the
conditional density matrix. Applying $K_{\mathrm y}$ to
Eq.~\eqref{eq:teleportidentity} gives
\begin{equation}
(K_{\mathrm y}\otimes I_3)
|\psi\rangle_1|\Phi^+\rangle_{23}
=\frac12\sum_{\beta\in\mathcal B}
\kappa_\beta  |\beta\rangle_{12}U_\beta|\psi\rangle_3.
\label{eq:filteredteleportation}
\end{equation}
Here $I_3$ denotes the identity operator on Bob's qubit $3$. A complete Bell
readout with result $\beta$ identifies $\beta$. Bob then applies
$U_\beta^\dagger$, recovering the input state exactly,
\begin{equation}
U_\beta^\dagger U_\beta|\psi\rangle=|\psi\rangle,
\end{equation}
and therefore
\begin{equation}
F_\beta=1
\label{eq:unitfidelity}
\end{equation}
for every accepted Bell-readout result.

The reduced state of qubits $1$ and $2$ before the
measurement is $\rho_1\otimes I_2/2$, where $\rho_1=|\psi\rangle\langle\psi|$.
The selected-branch probability is therefore independent of $c_0$ and $c_1$,
since $\operatorname{Tr}_2 E_{\mathrm y}=\operatorname{Tr}_2(P_\Phi+r^2P_\Psi)
=(1+r^2)I_1$. In the maximal-normalization convention,
\begin{align}
p_{\mathrm y}
&=\frac12\operatorname{Tr}\!\left[\rho_1\operatorname{Tr}_2 E_{\mathrm y}\right]
\nonumber\\
&=\frac{|A|^2+|B|^2}{2|A|^2}
=\frac{1+r^2}{2}
=\frac12\left[1+\frac{(1+\cos\theta)^2}{4}\right].
\label{eq:heraldingprobability}
\end{align}

\begin{figure}[tb]
\centering
\includegraphics[width=0.45\textwidth]{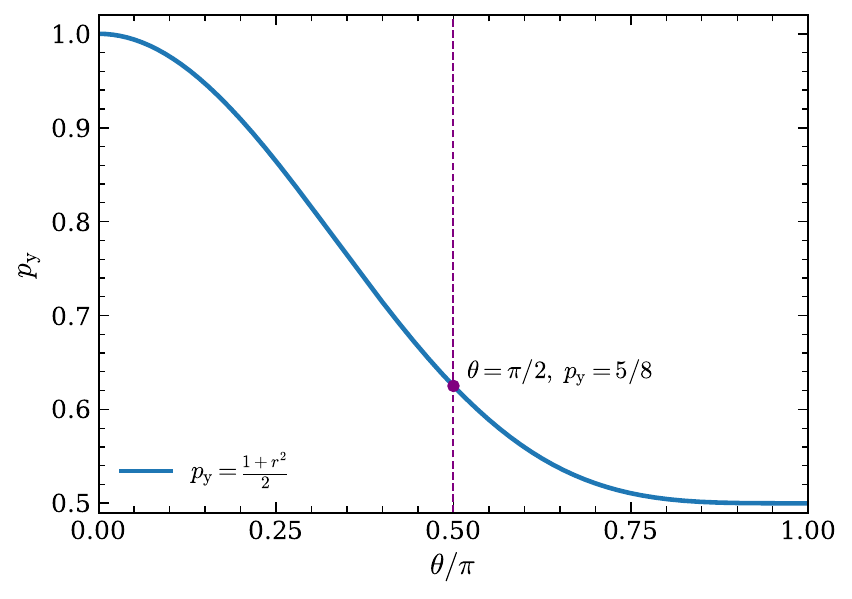}
\caption{
Angular dependence of the selected-branch probability
$p_{\mathrm y}=(1+r^2)/2$ in the maximal Kraus-normalization convention.
At $\theta=\pi/2$, one has $p_{\mathrm y}=5/8$.
The endpoint $\theta/\pi=1$ represents the normalized backward limit
$\theta\to\pi$.
}
\label{fig:heralding_probability}
\end{figure}

The angular dependence of the selected-branch probability $p_y$ is shown in 
Fig.~\ref{fig:heralding_probability}. 
In the maximal-normalization convention, $p_y$ decreases monotonically from unity at the 
forward angle to $1/2$ in the normalized backward limit. At $\theta=\pi/2$, one has $p_y=5/8$. 
Thus, although every accepted Bell-readout result yields unit conditional fidelity 
according to Eq.~\eqref{eq:unitfidelity}, 
the probability of reaching the selected Compton branch remains angle dependent.

This probability refers specifically to the maximal Kraus normalization in Eq.~\eqref{eq:Kyes}. 
If one instead uses $K=\alpha T$ with a smaller allowed $\alpha$, the selected-branch 
probability is rescaled by $\alpha^2|A|^2$, whereas the normalized conditional state and 
the unit fidelity in Eq.~\eqref{eq:unitfidelity} remain unchanged.

The two conclusions are therefore distinct. The incomplete binary measurement
alone never gives a quantum teleportation advantage. The selected Compton Kraus
branch can nevertheless serve as a heralded nonunitary element in an assisted
protocol when a complete Bell readout is subsequently supplied.

\section{Conclusion}\label{sec:conclusion}

We have presented a quantum circuit description of massless tree level
Compton scattering at fixed center-of-mass kinematics. In the helicity basis,
the scattering process is represented by a two-qubit transition map whose
structure becomes particularly transparent in the Bell basis. The Bell states
separate into two parity sectors, with the scattering map preserving each
sector and each Bell state as an eigenstate. This structure also admits an
exact quantum circuit representation consisting of unitary Bell-basis
transformations and a single nonunitary filtering operation.

After an appropriate Kraus normalization, the nonunitary part of the
scattering map defines an effective two-outcome measurement. This measurement
is sensitive to the Bell-parity sector but is completely insensitive to the
relative Bell label within each sector. It realizes an incomplete
Bell measurement. At a generic scattering angle, the measurement is 
unsharp, since a single outcome does not determine the Bell parity with
certainty. The amount of information contained in the binary measurement
record is consequently smaller than that of a complete Bell-state
measurement.

We have also distinguished two different teleportation scenarios. If only the
binary outcome of the incomplete measurement is retained, the resulting
teleportation fidelity does not exceed the classical threshold. In contrast,
the selected Compton branch preserves a one-to-one correspondence between the
incoming and outgoing Bell states at every nonsingular interior scattering
angle. When this branch is followed by a complete Bell-basis readout, the
unknown qubit can be recovered with unit conditional fidelity. 

These results clarify the quantum-information content of the Compton helicity
transition matrix and separate the information provided intrinsically by its
nonunitary structure from that supplied by an additional Bell measurement.
They provide a concrete realization of the quantum-circuit viewpoint and
illustrate how high energy scattering processes can be analyzed in terms of
quantum measurements, conditional operations, information flow, and their
possible role in quantum teleportation protocols.

\begin{acknowledgments}
WX was supported in part by the National Natural Science Foundation
of China under Grants Nos. 12375137 and 12005114. J.-C. Y. was supported
in part by the National Natural Science Foundation of China under 
Grants Nos. 12575106 and 12147214, the Basic Research Projects of Universities 
in Liaoning Province (Grant No.~LJ212510165024).

\end{acknowledgments}

\appendix

\section{External-state conventions and center-of-mass kinematics}
\label{app:conventions}

For completeness and reproducibility, we summarize here the external-state
conventions used in evaluating the helicity amplitudes. These conventions fix
the phases of the individual helicity matrix elements. Rephasing external
helicity states changes these phases but leaves singular values, probabilities,
and spin-summed observables unchanged.

We use natural units, $\hbar=c=1$, and the Minkowski metric
\begin{equation}
g^{\mu\nu}=\operatorname{diag}(1,-1,-1,-1).
\label{app:metric}
\end{equation}
The Dirac matrices are taken in the Dirac representation,
\begin{equation}
\gamma^0=
\begin{pmatrix}
I_2 & 0\\
0 & -I_2
\end{pmatrix},
\qquad
\gamma^i=
\begin{pmatrix}
0 & \sigma_i\\
-\sigma_i & 0
\end{pmatrix},
\qquad i=1,2,3,
\label{app:gammamatrices}
\end{equation}
where $I_2$ is the $2\times2$ identity matrix and $\sigma_i$ are the Pauli
matrices. We use
\begin{equation}
\slashed{a}\equiv \gamma_\mu a^\mu .
\end{equation}

\subsection{Center-of-mass kinematics}

In the massless limit, $m=0$, we work in the center-of-mass frame with beam
energy $E>0$. The scattering plane is chosen to be the $x$--$z$ plane, and
$\theta$ denotes the angle between the incoming and outgoing electron
momenta. The external four-momenta are
\begin{align}
p_1^\mu&=(E,0,0,E), \nonumber\\
p_2^\mu&=(E,0,0,-E), \nonumber\\
p_3^\mu&=(E,E\sin\theta,0,E\cos\theta), \nonumber\\
p_4^\mu&=(E,-E\sin\theta,0,-E\cos\theta).
\label{app:momenta}
\end{align}
For the external momenta in Eq.~\eqref{app:momenta}, we choose the angular
coordinates
\begin{equation}
(\vartheta_1,\varphi_1)=(0,0),\qquad
(\vartheta_2,\varphi_2)=(\pi,0),
\end{equation}
\begin{equation}
(\vartheta_3,\varphi_3)=(\theta,0),\qquad
(\vartheta_4,\varphi_4)=(\pi-\theta,\pi).
\end{equation}
These choices completely fix the external helicity phases used in the main
text.

The corresponding Mandelstam variables are
\begin{equation}
\begin{aligned}
s&=(p_1+p_2)^2=4E^2,\\
t&=(p_1-p_3)^2=-2E^2(1-\cos\theta),\\
u&=(p_1-p_4)^2=-2E^2(1+\cos\theta),
\end{aligned}
\label{app:stu}
\end{equation}
and therefore $s+t+u=0$.

\subsection{Electron helicity spinors}

For a unit three-vector
\begin{equation}
\widehat{\bm p}
=
(\sin\vartheta\cos\varphi,
 \sin\vartheta\sin\varphi,
 \cos\vartheta),
\end{equation}
we define the two-component helicity spinors by
\begin{align}
\chi_+(\vartheta,\varphi)
&=
\begin{pmatrix}
\cos(\vartheta/2)\\
e^{i\varphi}\sin(\vartheta/2)
\end{pmatrix},
\nonumber\\
\chi_-(\vartheta,\varphi)
&=
\begin{pmatrix}
-e^{-i\varphi}\sin(\vartheta/2)\\
\cos(\vartheta/2)
\end{pmatrix}.
\label{app:chi}
\end{align}
They satisfy
\begin{equation}
(\bm{\sigma}\cdot\widehat{\bm p})\chi_\lambda
=
\lambda\chi_\lambda,
\qquad
\lambda=\pm1 .
\end{equation}
The positive-energy massless Dirac spinors are chosen as
\begin{equation}
u(p,\lambda)
=
\sqrt{E}
\begin{pmatrix}
\chi_\lambda\\
\lambda\chi_\lambda
\end{pmatrix},
\qquad
\bar u(p,\lambda)=u^\dagger(p,\lambda)\gamma^0 .
\label{app:spinor}
\end{equation}
With this normalization,
\begin{equation}
u^\dagger(p,\lambda)u(p,\lambda)=2E,
\qquad
\slashed{p}\,u(p,\lambda)=0 .
\end{equation}

\subsection{Photon helicity polarization vectors}

For a photon propagating along
\begin{equation}
\widehat{\bm k}
=
(\sin\vartheta\cos\varphi,
 \sin\vartheta\sin\varphi,
 \cos\vartheta),
\end{equation}
we introduce the transverse unit vectors
\begin{align}
\bm e_\vartheta
&=
(\cos\vartheta\cos\varphi,
 \cos\vartheta\sin\varphi,
 -\sin\vartheta),
\nonumber\\
\bm e_\varphi
&=
(-\sin\varphi,\cos\varphi,0).
\label{app:transversebasis}
\end{align}
They satisfy
\begin{equation}
\bm e_\vartheta\cdot\bm e_\varphi=0,
\qquad
|\bm e_\vartheta|=|\bm e_\varphi|=1,
\qquad
\bm e_\vartheta\times\bm e_\varphi=\widehat{\bm k}.
\end{equation}
The photon helicity polarization vectors are chosen as
\begin{equation}
\epsilon^\mu(k,\lambda)
=
\frac{1}{\sqrt{2}}
\left(
0,\,
\bm e_\vartheta+i\lambda\bm e_\varphi
\right),
\qquad
\lambda=\pm1 .
\label{app:polarization}
\end{equation}
They obey
\begin{equation}
k\cdot\epsilon(k,\lambda)=0,
\qquad
\epsilon^*(k,\lambda)\cdot
\epsilon(k,\lambda')
=
-\delta_{\lambda\lambda'} .
\label{app:polarizationproperties}
\end{equation}
A change of the transverse reference axes corresponds to a rephasing of the
photon helicity states and therefore changes only convention-dependent phases
of individual helicity amplitudes, not the physical quantities considered in
the main text.

\bibliographystyle{utphys}
\bibliography{compton}

@article{Scherdin1991,
  title = {Low energy $e^+e^-$ scattering},
  author = {A. Scherdin, J. Reinhardt, W. Greiner and B. Muller},
  journal = {Rep. Prog. Phys.},
  volume = {54},
  pages = {1},
  year = {1991},
  doi = {10.1088/0034-4885/54/1/001},
  url = {https://doi.org/10.1088/0034-4885/54/1/001}
}

@article{Afik:2022kwm,
  author = {Afik, Y. and de Nova, J. R. M.},
  title = {Quantum information with top quarks in QCD},
  journal = {Quantum},
  volume = {6},
  year = {2022},
  pages = {820},
  doi = {10.22331/q-2022-09-29-820}
}

@article{Afik:2023dgh,
  author = {Afik, Y. and de Nova, J. R. M.},
  title = {Quantum discord and steering in top quarks at the LHC},
  journal = {Phys. Rev. Lett.},
  volume = {130},
  year = {2023},
  pages = {221801},
  doi = {10.1103/PhysRevLett.130.221801}
}

@article{QIHEP,
  author = {Afik, Y. and others},
  title = {Quantum information meets high-energy physics: {I}nput to the update of the {E}uropean strategy for particle physics},
  journal = {Eur. Phys. J. Plus},
  volume = {140},
  year = {2025},
  pages = {855},
  doi = {10.1140/epjp/s13360-025-06752-9}
}

@article{HZZ,
  author = {Aguilar-Saavedra, J. A. and Bernal, A. and Casas, J. A. and Moreno, J. M.},
  title = {Testing entanglement and Bell inequalities in H $\rightarrow$ ZZ},
  journal = {Phys. Rev. D},
  volume = {107},
  year = {2023},
  pages = {016012},
  doi = {10.1103/PhysRevD.107.016012}
}

@article{Araujo,
  author = {Araujo, J. B. and others},
  title = {Measuring QED cross sections via entanglement},
  journal = {Phys. Rev. D},
  volume = {100},
  year = {2019},
  pages = {105018},
  doi = {10.1103/PhysRevD.100.105018}
}

@article{Asenov:2025ueu,
  author = {Asenov, P. and Zhang, W. and Serafini, A.},
  title = {Quantum estimation in QED scattering},
  journal = {Phys. Rev. D},
  doi = {10.1103/b81h-td1t},
  url = {https://link.aps.org/doi/10.1103/b81h-td1t},
  volume = {113},
  year = {2026},
  number = {1},
  pages = {016015}
}

@misc{ATLAS:2026hye,
    author = "{ATLAS Collaboration}",
    title = "{Measurements of $Z$-boson pair entanglement in decays of Higgs bosons at the ATLAS experiment}",
    eprint = "2603.26463",
    archivePrefix = "arXiv",
    primaryClass = "hep-ex",
    year = "2026",
    url={https://arxiv.org/abs/2603.26463},
}

@article{Banerjee:2015mha,
    author = "Banerjee, Subhashish and Alok, Ashutosh Kumar and Srikanth, R. and Hiesmayr, Beatrix C.",
    title = "{A quantum information theoretic analysis of three flavor neutrino oscillations}",
    journal = "Eur. Phys. J. C",
    volume = "75",
    number = "10",
    pages = "487",
    year = "2015"
}

@article{Barr:2024djo,
    author = "Barr, Alan J. and Fabbrichesi, Marco and Floreanini, Roberto and Gabrielli, Emidio and Marzola, Luca",
    title = "{Quantum entanglement and Bell inequality violation at colliders}",
    doi = "10.1016/j.ppnp.2024.104134",
    journal = "Prog. Part. Nucl. Phys.",
    volume = "139",
    pages = "104134",
    year = "2024"
}

@article{Barr:2021zcp,
    author = "Barr, Alan J.",
    title = "{Testing Bell inequalities in Higgs boson decays}",
    doi = "10.1016/j.physletb.2021.136866",
    journal = "Phys. Lett. B",
    volume = "825",
    pages = "136866",
    year = "2022"
}

@article{Beane:2018oxh,
  author = {Beane, S. R. and Kaplan, D. B. and Klco, N. and Savage, M. J.},
  title = {Entanglement Suppression and Emergent Symmetries of Strong Interactions},
  journal = {Phys. Rev. Lett.},
  volume = {122},
  year = {2019},
  pages = {102001},
  doi = {10.1103/PhysRevLett.122.102001}
}

@article{Bertlmann:2001ea,
    author = "Bertlmann, R. A. and Grimus, W. and Hiesmayr, B. C.",
    title = "{Bell inequality and CP violation in the neutral kaon system}",
    doi = "10.1016/S0375-9601(01)00577-1",
    journal = "Phys. Lett. A",
    volume = "289",
    pages = "21--26",
    year = "2001"
}

@incollection{Bertlmann2006,
    author    = {Bertlmann, R. A.},
    title     = {{Entanglement, Bell inequalities and decoherence in particle physics}},
    booktitle = {Quantum coherence: from quarks to solids},
    editor    = {P{\"o}tz, Walter and Hohenester, Ulrich and Fabian, Jaroslav},
    series    = {Lecture Notes in Physics},
    volume    = {689},
    pages     = {1--45},
    year      = {2006},
    publisher = {Springer},
    doi       = {10.1007/11398448_1}
}

@article{Bittencourt:2022tcl,
    author = "Bittencourt, V. A. S. V. and Blasone, M. and De Siena, S. and Matrella, C.",
    title = "{Complete complementarity relations for quantum correlations in neutrino oscillations}",
    doi = "10.1140/epjc/s10052-022-10508-5",
    journal = "Eur. Phys. J. C",
    volume = "82",
    number = "6",
    pages = "566",
    year = "2022"
}

@article{Bittencourt:2023asd,
    author = "Bittencourt, V. A. S. V. and Blasone, Massimo and De Siena, Silvio and Matrella, Cristina",
    title = "{Quantifying quantumness in three-flavor neutrino oscillations}",
    doi = "10.1140/epjc/s10052-024-12631-x",
    journal = "Eur. Phys. J. C",
    volume = "84",
    number = "3",
    pages = "301",
    year = "2024"
}

@article{BlasoneSE2,
  author = {Blasone, M. and De Siena, S. and Lambiase, G. and Matrella, C. and Micciola, B.},
  title = {Complete complementarity relations in tree level QED processes},
  journal = {Phys. Rev. D},
  volume = {111},
  year = {2025},
  pages = {016007},
  doi = {10.1103/PhysRevD.111.016007}
}

@article{NoiChaos,
  author = {Blasone, M. and De Siena, S. and Lambiase, G. and Matrella, C. and Micciola, B.},
  title = {Entanglement dynamics in QED processes},
  journal = {Chaos, Solitons \& Fractals},
  volume = {195},
  year = {2025},
  pages = {116305},
  doi = {10.1016/j.chaos.2025.116305}
}

@article{Blasone:2024dud,
  author = {Blasone, M. and Lambiase, G. and Micciola, B.},
  title = {Entanglement distribution in Bhabha scattering with entangled spectator particle},
  journal = {Phys. Rev. D},
  volume = {109},
  year = {2024},
  pages = {096022},
  doi = {10.1103/PhysRevD.109.096022}
}

@article{Blasone:2021cau,
    author = "Blasone, Massimo and De Siena, Silvio and Matrella, Cristina",
    title = "{Wave packet approach to quantum correlations in neutrino oscillations}",
    doi = "10.1140/epjc/s10052-021-09471-4",
    journal = "Eur. Phys. J. C",
    volume = "81",
    number = "7",
    pages = "660",
    year = "2021"
}

@article{Blasone:2023qqf,
    author = "Blasone, Massimo and De Siena, Silvio and Matrella, Cristina",
    title = "{Complete complemetarity relations and quantifiers of quantum correlations in neutrino oscillations}",
    doi = "10.22323/1.436.0044",
    journal = "PoS",
    volume = "CORFU2022",
    pages = "044",
    year = "2023"
}

@article{Blasone:2007vw,
    author = "Blasone, Massimo and Dell'Anno, Fabio and De Siena, Silvio and Illuminati, Fabrizio",
    title = "{Entanglement in neutrino oscillations}",
    doi = "10.1209/0295-5075/85/50002",
    journal = "EPL",
    volume = "85",
    pages = "50002",
    year = "2009"
}

@misc{Cao:2026ljz,
    author = "Cao, Zan and Song, Meng-Long and Song, Xue-Ke and Ye, Liu and Wang, Dong",
    title = "{Dynamical redistribution of quantum resources in tree-level Bhabha scattering}",
    eprint = "2607.20264",
    archivePrefix = "arXiv",
    primaryClass = "quant-ph",
    month = "7",
    year = "2026"
}

@article{Cao:2026mza,
    author = "Cao, Zan and Song, Meng-Long and Song, Xue-Ke and Ye, Liu and Wang, Dong",
    title = "{Genuine tripartite entanglement in Bhabha scattering with an entangled spectator particle}",
    doi = "10.1103/1885-yhgf",
    journal = "Phys. Rev. D",
    volume = "113",
    number = "10",
    pages = "105019",
    year = "2026"
}

@article{Carena:2025wyh,
    author = "Carena, Marcela and Coloretti, Guglielmo and Liu, Wanqiang and Littmann, Mira and Low, Ian and Wagner, Carlos E. M.",
    title = "{Entanglement maximization and mirror symmetry in two-Higgs-doublet models}",
    doi = "10.1007/JHEP08(2025)016",
    journal = "J. High Energ. Phys.",
    volume = "08",
    pages = "016",
    year = "2025"
}

@article{Cervera-Lierta:2017tdt,
  author = {Cervera-Lierta, A. and Latorre, J. I. and Rojo, J. and Rottoli, L.},
  title = {Maximal Entanglement in High Energy Physics},
  journal = {SciPost Phys.},
  volume = {3},
  year = {2017},
  pages = {036},
  doi = {10.21468/SciPostPhys.3.5.036}
}

@article{Chang:2024wrx,
    author = "Chang, Spencer and Jacobo, Gabriel",
    title = "{Consequences of minimal entanglement in bosonic field theories}",
    doi = "10.1103/PhysRevD.110.096020",
    journal = "Phys. Rev. D",
    volume = "110",
    number = "9",
    pages = "096020",
    year = "2024"
}

@misc{Clarke2026,
      title = {Bell Test of Photons from Electron-Positron Annihilation via {POVM}-based {C}ompton Polarimetry}, 
      author = {Jack Clarke and Preslav Asenov and Jesse Smeets and Jia-Shian Wang and David B. Cassidy and Alessio Serafini},
      year = {2026},
      eprint = {2604.25034},
      archivePrefix = {arXiv},
      primaryClass = {quant-ph}
}

@article{DiDomenico:1995ky,
    author = "Di Domenico, A.",
    title = "{Testing quantum mechanics in the neutral kaon system at a Phi factory}",
    doi = "10.1016/0550-3213(95)00283-X",
    journal = "Nucl. Phys. B",
    volume = "450",
    pages = "293--324",
    year = "1995"
}

@book{DiDomenico:2007zza,
    author = "Di Domenico, Antonio",
    title = "{Handbook on neutral kaon interferometry at a $\Phi$-factory}",
    publisher = "INFN",
    address = "Frascati, Italy",
    series = "Frascati physics series",
    volume = "43",
    year = "2007"
}

@article{Dixit:2018gjc,
    author = "Dixit, Khushboo and Naikoo, Javid and Banerjee, Subhashish and Kumar Alok, Ashutosh",
    title = "{Study of coherence and mixedness in meson and neutrino systems}",
    doi = "10.1140/epjc/s10052-019-6609-7",
    journal = "Eur. Phys. J. C",
    volume = "79",
    number = "2",
    pages = "96",
    year = "2019"
}

@article{Duch:2024pwm,
    author = "Duch, Mateusz and Strumia, Alessandro and Titov, Arsenii",
    title = "{New physics in spin entanglement}",
    doi = "10.1140/epjc/s10052-025-13836-4",
    journal = "Eur. Phys. J. C",
    volume = "85",
    number = "2",
    pages = "151",
    year = "2025"
}

@article{Fabbrichesi:2021npl,
    author = "Fabbrichesi, M. and Floreanini, R. and Panizzo, G.",
    title = "{Testing Bell Inequalities at the LHC with Top-Quark Pairs}",
   doi = "10.1103/PhysRevLett.127.161801",
    journal = "Phys. Rev. Lett.",
    volume = "127",
    number = "16",
    pages = "161801",
    year = "2021"
}

@article{Fan,
  author = {Fan, J. and Deng, G. M. and Ren, X. J.},
  title = {Entanglement entropy and monotones in scattering process},
  journal = {Phys. Rev. D},
  volume = {104},
  year = {2021},
  pages = {116021},
  doi = {10.1103/PhysRevD.104.116021}
}

@article{Fan:2017mth,
  author = {Fan, J. and Li, X.},
  title = {Relativistic effect of entanglement in fermion-fermion scattering},
  journal = {Phys. Rev. D},
  volume = {97},
  year = {2018},
  pages = {016011},
  doi = {10.1103/PhysRevD.97.016011}
}

@article{Serafini,
  author = {Fedida, S. and Serafini, A.},
  title = {Tree-level entanglement in Quantum Electrodynamics},
  journal = {Phys. Rev. D},
  volume = {107},
  year = {2023},
  pages = {116007},
  doi = {10.1103/PhysRevD.107.116007}
}

@article{Fedida:2024dwc,
    author = "Fedida, Samuel and Mazumdar, Anupam and Bose, Sougato and Serafini, Alessio",
    title = "{Entanglement entropy in scalar quantum electrodynamics}",
    eprint = "2401.10332",
    archivePrefix = "arXiv",
    primaryClass = "hep-th",
    doi = "10.1103/PhysRevD.109.065028",
    journal = "Phys. Rev. D",
    volume = "109",
    number = "6",
    pages = "065028",
    year = "2024"
}

@article{Fonseca:2021uhd,
  author = {Fonseca, J. D. and others},
  title = {Entanglement and scattering in quantum electrodynamics: S-matrix information from an entangled spectator particle},
  journal = {Phys. Rev. D},
  volume = {106},
  year = {2022},
  pages = {056015},
  doi = {10.1103/PhysRevD.106.056015}
}

@article{Formaggio:2016cuh,
    author = "Formaggio, J. A. and Kaiser, D. I. and Murskyj, M. M. and Weiss, T. E.",
    title = "{Violation of the Leggett-Garg Inequality in Neutrino Oscillations}",
    doi = "10.1103/PhysRevLett.117.050402",
    journal = "Phys. Rev. Lett.",
    volume = "117",
    number = "5",
    pages = "050402",
    year = "2016"
}

@article{Gao:2025kdi,
    author = "Gao, Leyun and Ruzi, Alim and Li, Qite and Zhou, Chen and Li, Qiang",
    title = "{Testing entanglement between free-traveling electron-positron pairs}",
    primaryClass = "hep-ph",
    doi = "10.1103/m9z9-1jpd",
    journal = "Phys. Rev. D",
    volume = "111",
    number = "11",
    pages = "116018",
    year = "2025"
}

@article{Goebel1993,
author={G{\"o}bel, R. and Arnold, W. and Frommhold, Th. and Stock, R. and Weber, Th. and Kneissl, U. and Steiper, F. and Kozhuharov, C. and Kienle, P.},
title={Bhabha to M{\o}ller cross section ratios in the energy range 2.15--2.45 MeV},
journal={Z. Phys. A: Hadron Nucl.},
year={1993},
volume={345},
number={1},
pages={79--88},
doi={10.1007/BF01290343},
url={https://doi.org/10.1007/BF01290343}
}

@article{LHC,
  author = {Han, T. and Low, M. and McGinnis, N. and Su, S.},
  title = {Measuring quantum discord at the LHC},
  journal = {J. High Energ. Phys.},
  volume = {2025},
  year = {2025},
  pages = {81},
  doi = {10.1007/JHEP05(2025)081}
}

@article{CMS,
   title = {Measurements of polarization and spin correlation and observation of entanglement in top quark pairs using $\mathrm{lepton}+\text{jets}$ events from proton-proton collisions at $\sqrt{s}=13\text{ }\text{ }\mathrm{TeV}$},
    author = {Hayrapetyan, A. and others},
    collaboration = {{CMS Collaboration}},
  journal = {Phys. Rev. D},
  volume = {110},
  pages = {112016},
  year = {2024},
  doi = {10.1103/PhysRevD.110.112016},
}

@article{JacobWick,
  author = {Jacob, M. and Wick, G. C.},
  title = {On the general theory of collisions for particles with spin},
  journal = {Ann. Phys.},
  volume = {281},
  year = {2000},
  pages = {774},
  doi = {10.1006/aphy.2000.6022},
  note = {reprint of Nucl. Phys. 7, 407 (1964)}
}

@article{Jaloum:2025bkx,
    author = "Jaloum, Elhabib and Amazioug, Mohamed",
    title = "{Controlling the dynamical evolution of quantum coherence and quantum correlations in e+e-{\textrightarrow}{\ensuremath{\Lambda}}{\ensuremath{\Lambda}}{\textasciimacron} processes at BESIII}",
    doi = "10.1103/kxw9-wdth",
    journal = "Phys. Rev. D",
    volume = "113",
    number = "1",
    pages = "016024",
    year = "2026"
}

@article{Kowalska:2024kbs,
    author = "Kowalska, Kamila and Sessolo, Enrico Maria",
    title = "{Entanglement in flavored scalar scattering}",
    doi = "10.1007/JHEP07(2024)156",
    journal = "J. High Energy Phys.",
    volume = "2024",
    pages = "156",
    year = "2024"
}

@article{Xie:2026ixr,
    author = "Xie, Wei and Yang, Ji-Chong",
    title = "{A collider as a quantum computer}",
    eprint = "2605.06907",
    archivePrefix = "arXiv",
    primaryClass = "hep-ph",
    month = "5",
    year = "2026"
}

@article{Cervera-Lierta:2019juk,
    author = "Cervera-Lierta, Alba and Latorre, Jos{\'e} Ignacio and Goyeneche, Dardo",
    title = "{Quantum circuits for maximally entangled states}",
    eprint = "1904.07955",
    archivePrefix = "arXiv",
    primaryClass = "quant-ph",
    doi = "10.1103/PhysRevA.100.022342",
    journal = "Phys. Rev. A",
    volume = "100",
    number = "2",
    pages = "022342",
    year = "2019"
}

@book{NielsenChuang,
  author    = {Nielsen, Michael A. and Chuang, Isaac L.},
  title     = {Quantum Computation and Quantum Information},
  publisher = {Cambridge University Press},
  address   = {Cambridge},
  year      = {2010}
}

\end{document}